\pdfoutput=1
\documentclass{article}

\usepackage{arxiv}
\usepackage[utf8]{inputenc} 
\usepackage[T1]{fontenc}    
\usepackage{url}            
\usepackage{amsfonts}       
\usepackage{nicefrac}       
\usepackage{microtype}      
\usepackage{lipsum}
\usepackage{graphicx}
\usepackage{epstopdf}
\usepackage{hyperref}
\usepackage[section]{placeins}

\title{Trends in Cuban research output: publications and patents}

\author{
  Yasset Perez-Riverol \\
  Independent scholar\\
  Cambridge, UK \\
  \texttt{ypriverol@gmail.com} \\
}

\begin{document}
\maketitle

\begin{abstract}
Cuban science and technology are known for important achievements, particularly in human health care and biotechnology. During the second half of XX century, the country developed a system of scientific institutions to address and solve major economical, cultural, social and health problems. However, the economic crisis faced by the island during the last three decades has had a major impact in Cuban scientific research. In addition to decreased investment, the emigration of thousands of young as well as senior scientists to other countries have had a major impact in Cuban research output. To date, no systematic analysis regarding scientific publications, citations, or patents granted to Cuban authors during this period, are available. Here, an analysis of Cuban scientific production since 1970, with an especial focus on the last three decades (1990 - 2019), is provided. All national metrics are compared with other countries, emphasizing those from Latin America. Preliminary results show that Cuban scientific publications are increased at a lower rate (two-fold) compare with several Latin American countries (five-fold average). In addition, since 2014 the annual number of Cuban scientific publications is decreasing. Finally, the analysis shows that most young Cuban authors with the higher index of citations (1990-2019) are working abroad. All the data and the code related to this study are open and can be found in \href{https://github.com/ypriverol/cubascience}{GitHub}.
\end{abstract}


\section{Introduction}

During the last decades, Latin American (LA) countries have significantly increased their research investment and consequently their scientific output \cite{VanNoorden2014, Bajak2018, Bajak2019}. Countries such as Brazil, Chile, and Colombia have experienced unprecedented growth in the number of university facilities, research institutions, scientific publications, citations and patents granted. In Latin America, Cuba is an example of how to translate basic science from universities and research institutes into a strong health care system \cite{2009}. The production of the antimeningococcal vaccine (VA-MENGOC-GOC(R)) a completely novel vaccine for meningitis type B in the ’80s and it's application in the general population reduced the number of cases to almost eradication. More recently, Cuban biotechnology developed and obtained several biologics such as Heperprot-P for the treatment of diabetic ulcers and CimaVax-EGF, a cutting edge monoclonal antibody for lung cancer.

Most Cuban scientific production is related to biomedical and biotechnology research, and is produced by the University of Havana and the so called Scientific Pole, also located in the capital of the island. By 2006, this complex system of I+D centers comprises more than 52 institutions with up to 7000 researchers and roughly 900 patents submitted. The Pole also has several collaborators and subordinated institutions located in different parts of the country. Historical data shows that, unlike biotechnology and biomedical research, other fields such as agriculture, telecommunications or economical sciences have not experienced the same booster on basic or applied research in Cuba. Similarly, scientific production is not homogeneous across the country, with significant variations among universities and research institutions from different regions or provinces.

Cuban scientific output has been previously studied by multiple authors \cite{arencibia2010challenges, dorta2016colaboracion, azevedo2019wbopendata,zacca2015patrones} with the caveat of being focused on specific fields, or having only publications as the main scientific output. Measuring the strength and quality of research is a complex task. However, multiple metrics are now available to evaluate and improve the quality of scientific research and to help policymakers, universities as well as governments to quantify the impact of their scientific production. Some of the most relevant metrics are, (i) the number of scientific publications indexed in  major databases (e.g. Scopus, Thomson Reuter’s Science Citation); (ii) the number of citations to these publications; (iii) number of submitted and approved patents. These indicators represent a good starting point to evaluate the current state and performance of research and scientific production in a given country. 

In this study, public data from multiple sources were analised to quantify the scientific output of Cuban researchers since 1970, with special focus in the last three decades (1990-2019). Databases such as Web of Science, SCImago Journal, Country Rank and the World Bank Open Data were used to compile the number of publications and patents of Cuban institutions and researchers. Other LA countries (e.g. Colombia, Brazil, Mexico) were included to evaluate recent patterns and trends within the region. Unlike previous studies, the Cuban scientific production was further analyzed by province, research institutions, and universities. Finally, a list with the 100 most cited Cuban authors living abroad and whiting the country was curated using Google Scholar and ResearchGate. The h-index, citations, and number of manuscripts were studied for both groups of authors to highlight the impact of immigration in Cuban science. A \href{https://ypriverol.github.io/cubascience/}{Jupyter notebook} and \href{https://github.com/ypriverol/cubascience}{GitHub repository} made the present data public for scientists and institutions for future research.

\section{Methods}

The data from scientific publications were collected from  \href{https://www.scimagojr.com/}{SCImago Journal and Country Rank}, and \href{https://apps.webofknowledge.com}{Web of Science}, data retrieved February 2020. The patent's information was downloaded from the \href{https://data.worldbank.org/}{World Bank Open Data}. We only consider patents from Cuban resident applicants. All data can be openly accessed from the following \href{https://github.com/ypriverol/cubascience}{GitHub repository}. Also, a list of Cuban authors with over one thousand citations in Google Scholar was mannually curated. The analysis was performed using Jupyter notebook and Python version 3.0. A \href{https://ypriverol.github.io/cubascience/}{web page} was created for full access to all the plots, analysis, and data. The deployment of the web page is automatic, then external users can contribute to the data files and python notebook and the results will be seen on the web page automatically. 

\section{Results and Discussion}

\subsection{Scientific publications and citations}
 
\begin{figure}[ht]
\begin{tabular}{cc}
  \includegraphics[width=8cm]{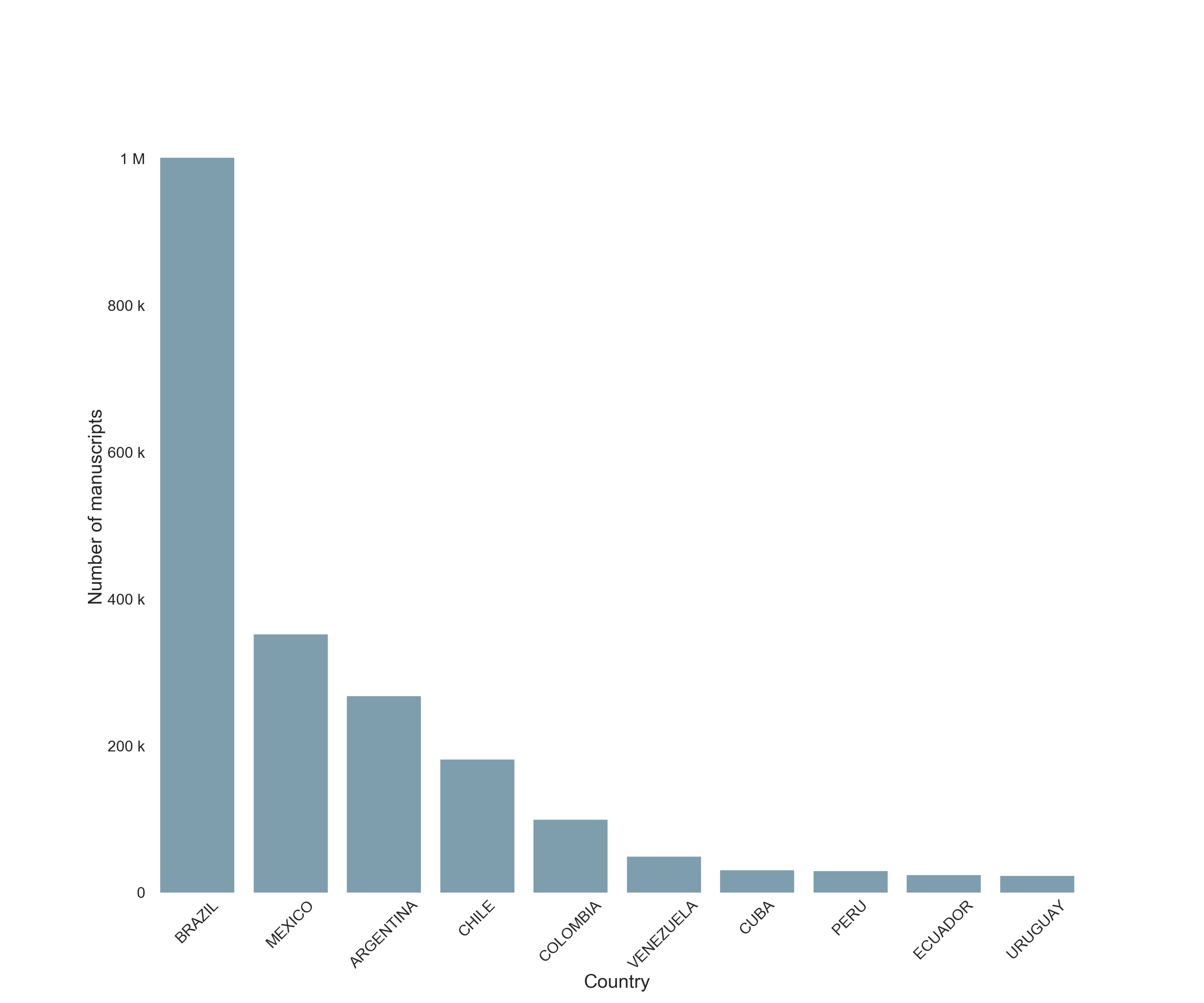} &   \includegraphics[width=8cm]{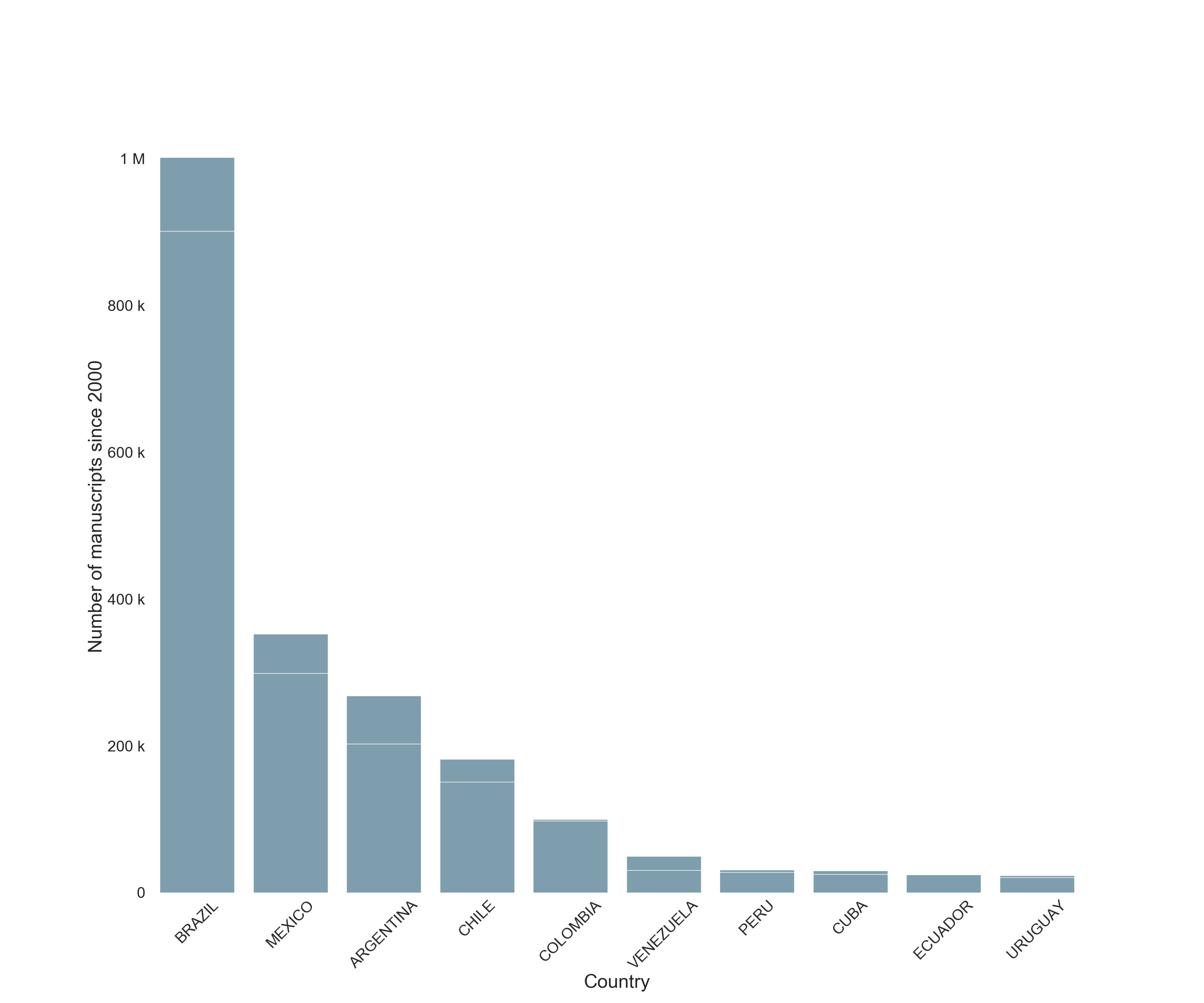} \\
  (a) Number of publications per country since 1970 & (b) Number of publications per country since 2020 \\[6pt]
  \includegraphics[width=8cm]{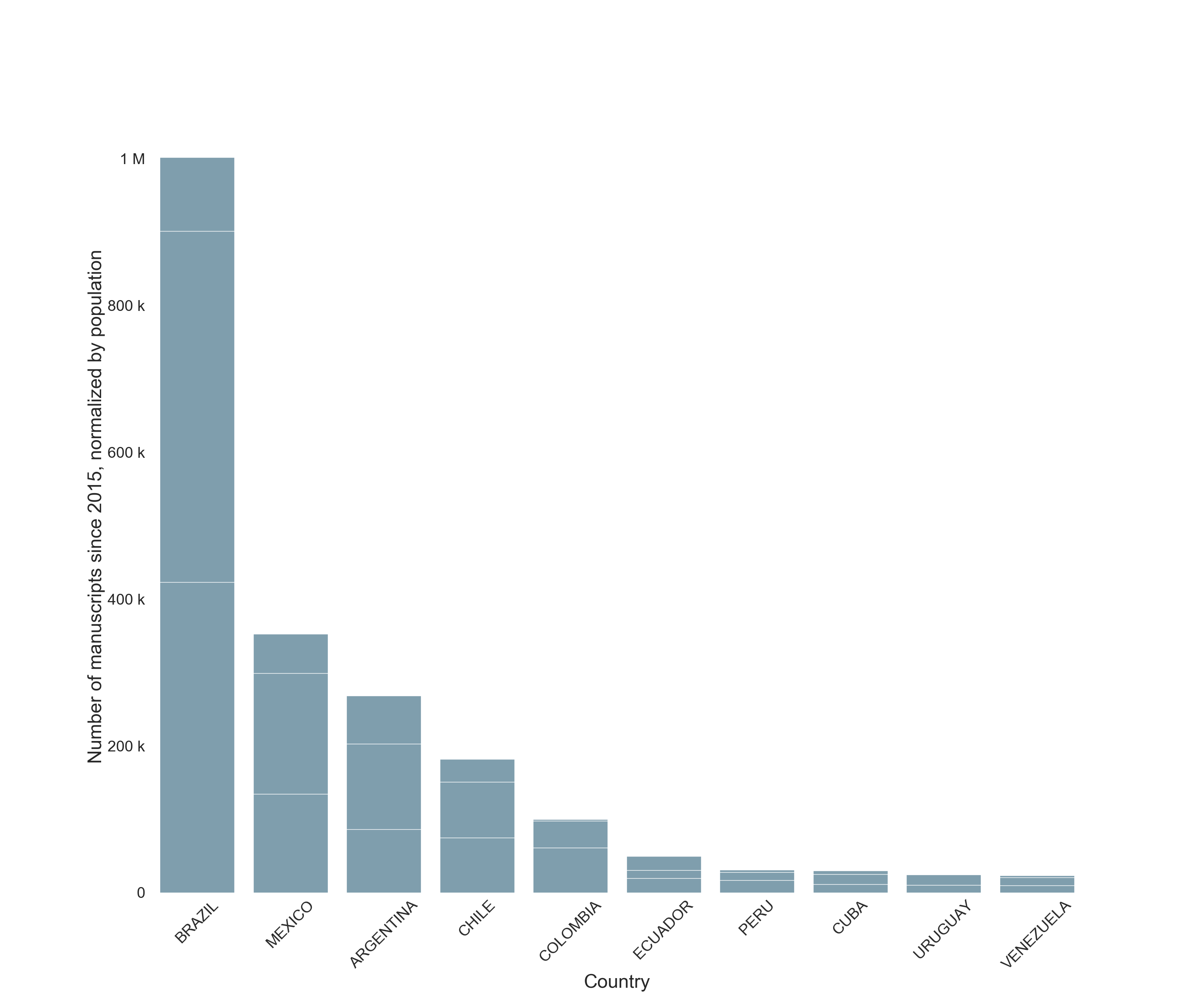} \\
  (c) Number of publications per country since 2015 \\[6pt]
\end{tabular}
\caption{Number of publications per country in Latin America (a) 1970 - 2020, (b) 2000 - 2020, (c) 2015 - 2020}
\label{fig:fig1}
\end{figure}

From 19170-2019, over 30’000 scientific publications from Cuba, were indexed on the Web of Science database (Figure \ref{fig:fig1}). During the same period, Brazil, Mexico, Argentina, Chile, Colombia, and Venezuela published 350’000, 260’000, 180'000, 99'000, and 50’000 manuscripts, respectively. In Latin America, Cuba ranks seven in the list of top countries by the total number of publications from 1970 - 2020 (Figure \ref{fig:fig1}a) as well as number in the per capita of publications normalized by population size (data not shown). From 2000-2020, the country ranks eight in the region as Peru showed higher levels of manuscripts publication (Figure \ref{fig:fig1} b). Despite being overcome also by Ecuador, during the last five years Cuba still ranked eight in the region as Venezuela experienced a drastic decrease it corresponding scientific output. The significant increasing on research production of several countries in the region combined with the partial decrease of Cuban production represent major causes for this behaviour. Confraria and Vargas \cite{Confraria2017} found that the LA scientific publications started to increase at a higher rate since 1993, thus revealing a trend for convergence with the world-leading regions. This increase has been mainly driven by Brazil, most notably in subject areas such as Agricultural, and Plant and Animal Sciences. 

\begin{figure}[ht]
\begin{tabular}{cc}
  \includegraphics[width=8cm]{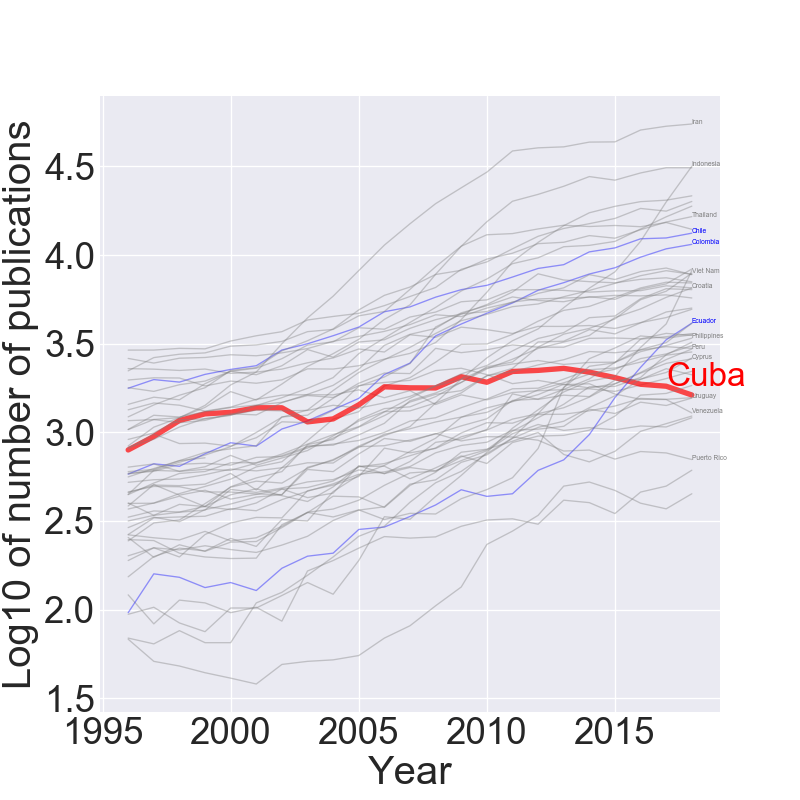} &   \includegraphics[width=8cm]{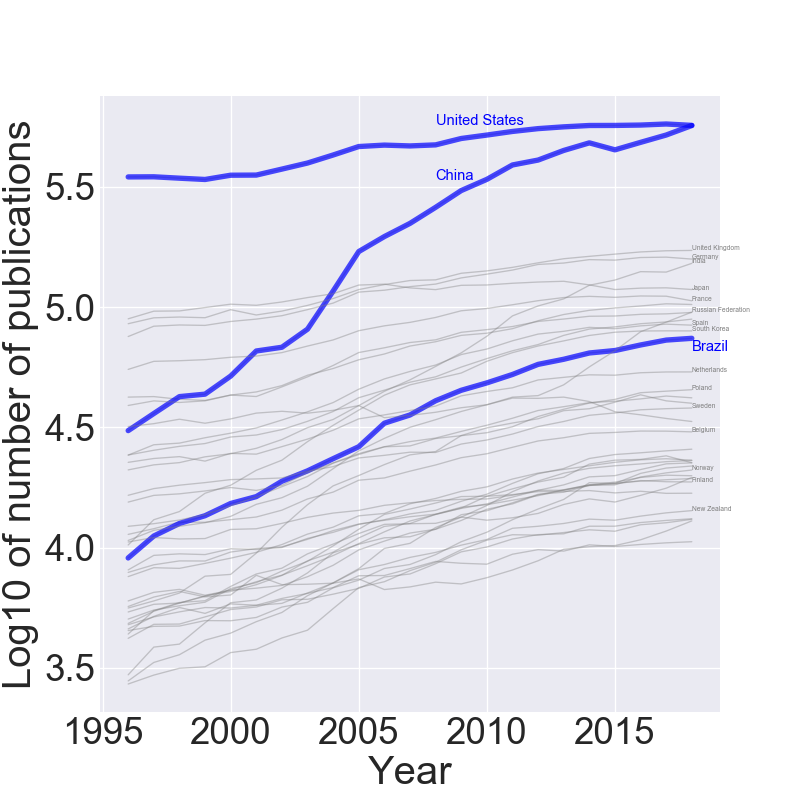} \\
  (a) Countries with lower number of publications & (b) Countries with higher number of publications \\[6pt]
  \includegraphics[width=8cm]{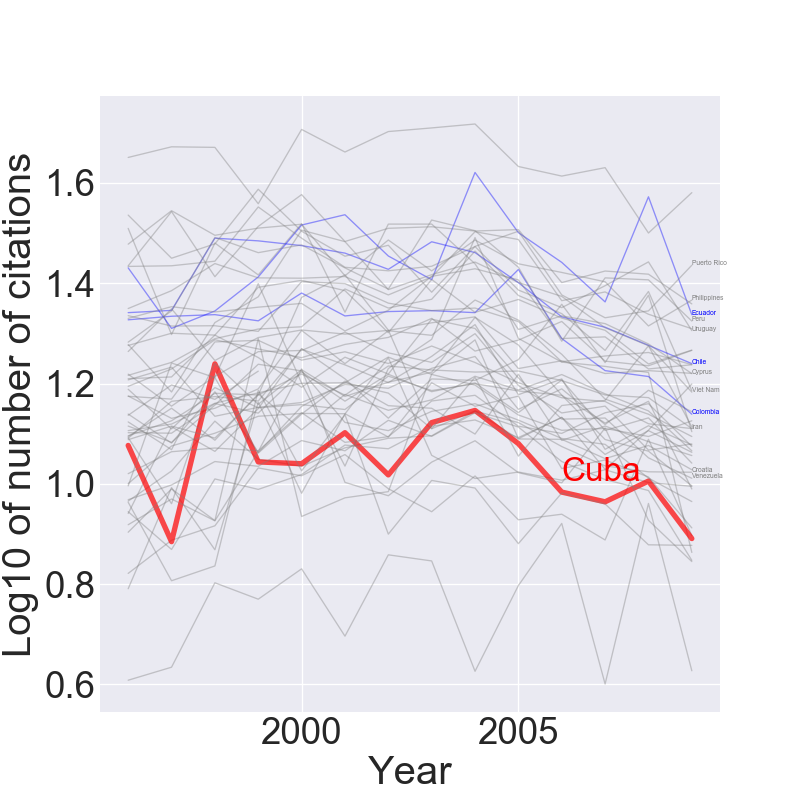} & \includegraphics[width=8cm]{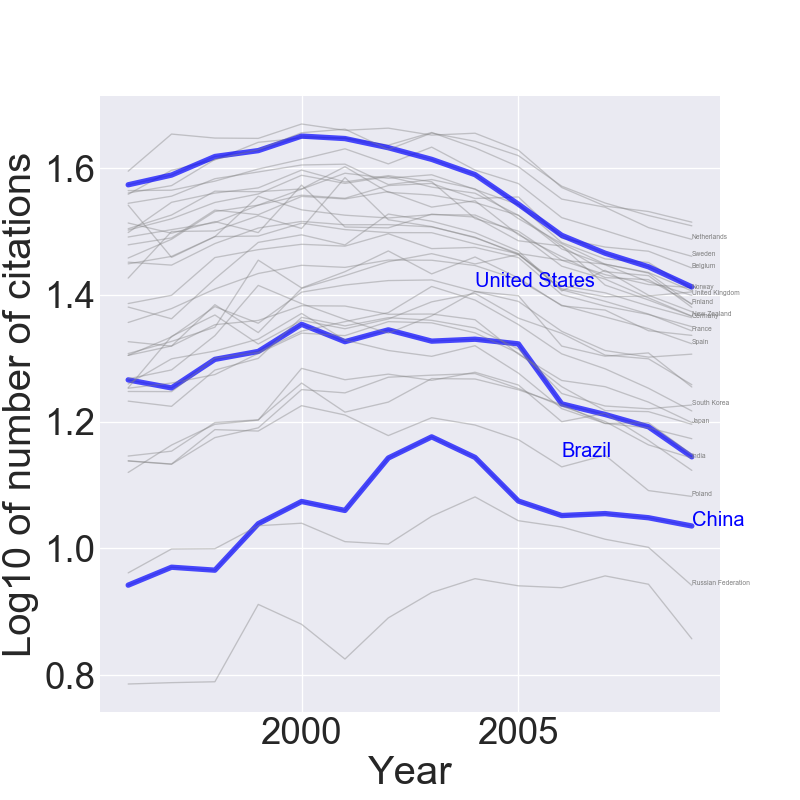} \\
  (c) Countries with lower number of citations & (d) Countries with higher number of citations \\[6pt]
\end{tabular}
\caption{Countries with lower and higher number of publications (a,b) and citations (c,d) per year (Time interval between 1995 - 2018)}
\label{fig:fig2}
\end{figure}

The total amount of Cuban scientific publications increased 2-fold from 1996-2018 (Figure \ref{fig:fig2} a), while other LA countries such as Ecuador, Chile, and Colombia increased 43, 7.5, and 19-fold, respectively. Remarkably, the number of Cuban research publications is decreasing since 2014. In 2013, 2468 scientific articles were published, while in 2014 (2332), 2015 (2182), 2016 (2026), 2017 (1990) and 2018 (1806). Countries with a higher number of publications (Figure \ref{fig:fig2}b) are also increasing the number of publications every year. This trend, along with the overall lower increase of Cuban scientific production (1996 to present) as compared to other countries in the region could be partially explained by the the fact that around 78\% of the total publications of the country (2000-2016) were published in national journals\cite{galban2019measuring}, not indexed on the Web of Science. As noted, the analysis was performed using only manuscripts published on international, peer-reviewed journals indexed on this database. Manuscripts published in Cuban national journals (most written in Spanish) were excluded in the present analysis as they are difficult to compare with other regions, and countries. 

The number of citations per manuscript of Cuban publications (produced by Cuban Institutions) also show a decrease since 2004 (Figure \ref{fig:fig2} c, d). The number of citations per manuscript of Cuba is lower than Chile, Colombia, or Uruguay (Figure \ref{fig:fig2} c). The low number of citations can be related to the nature or subject of the published manuscript. In general terms, applied sciences and translational research (e.g. description of a company product) which represents the core of Cuban scientific activity received fewer citations than basic sciences (e.g. explanation of a biological mechanism). In contrast, it would be expected an increase in the number of submitted and approved patents which is highly associated to reinforcement of applied sciences.    

\subsection{Applied sciences through patents}

\begin{figure}[ht]
\begin{tabular}{cc}
  \includegraphics[width=8cm]{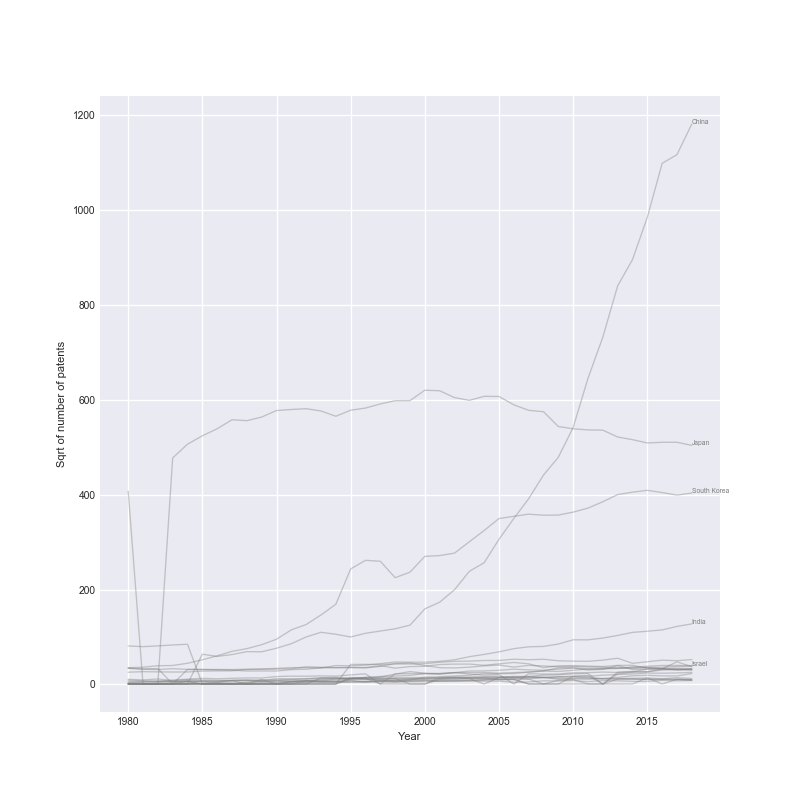} &   \includegraphics[width=8cm]{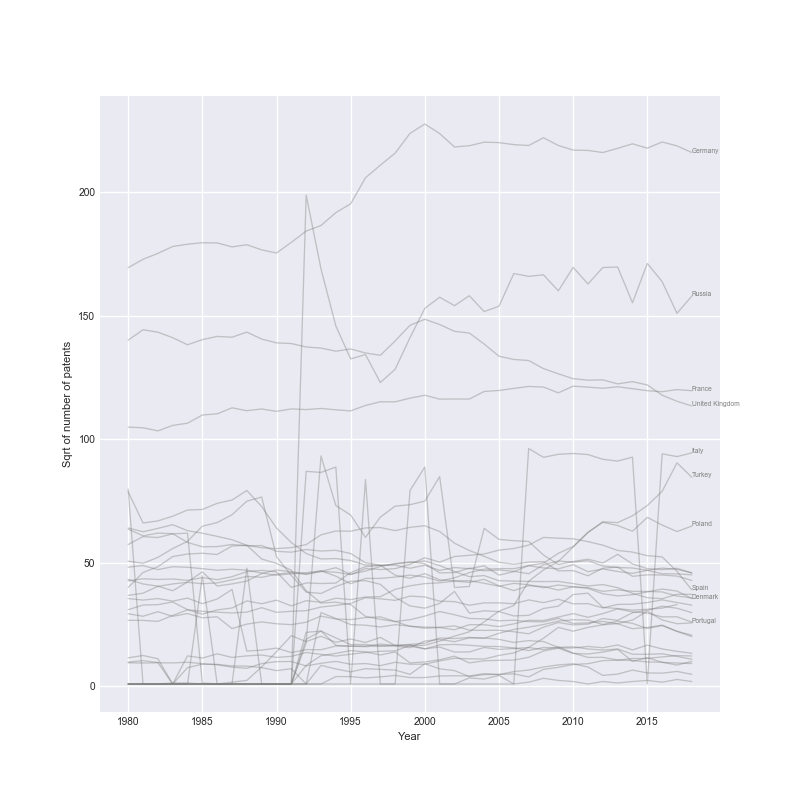} \\
  (a) Asia & (b) Europe \\[6pt]
  \includegraphics[width=8cm]{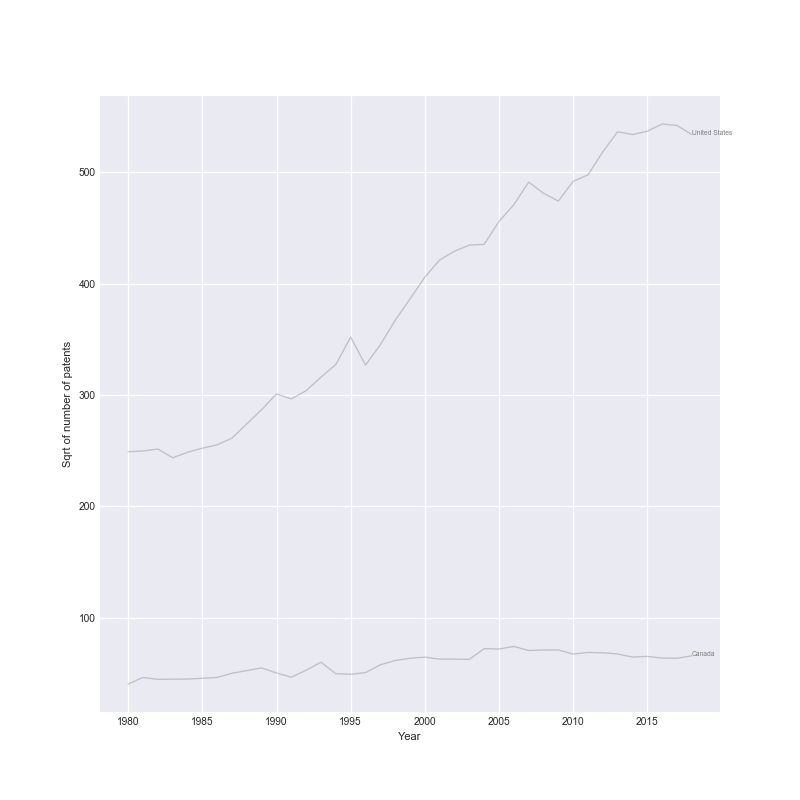} &   \includegraphics[width=8cm]{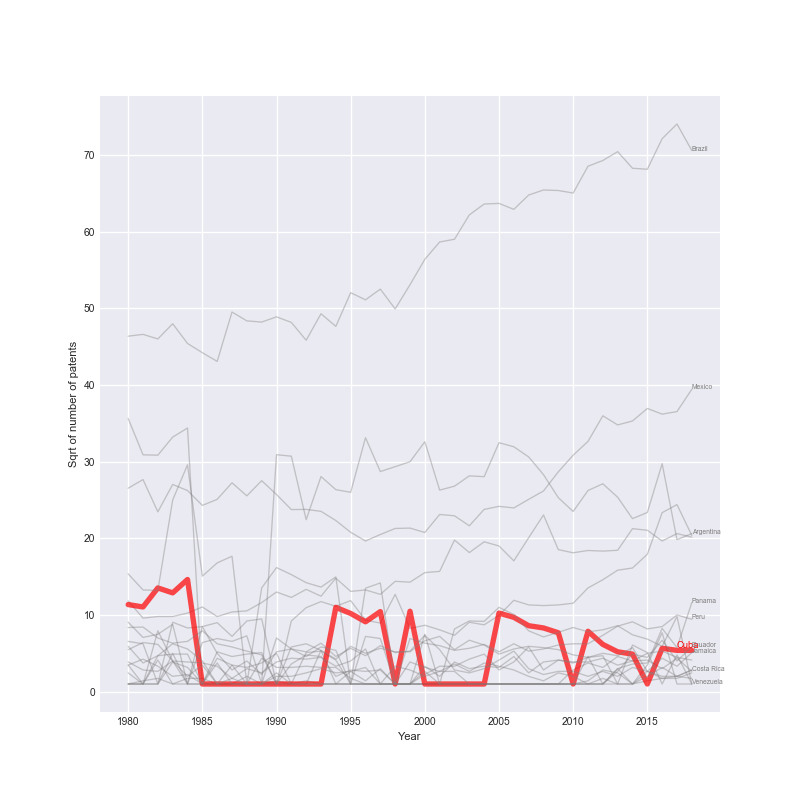} \\
  (c) North America  & (d) South America \\[6pt]
\end{tabular}
\caption{Number of patents (square root scale) by year, country and continent. In the plot, 0 values should be considered as no information is present for the specific year.}
\label{fig:fig3}
\end{figure}

To analyze the recent developments of translational and applied science in Cuba, the number of patents granted to Cuban residents from 1980 to 2018, were studied (Figure \ref{fig:fig3}). A patent is a form of intellectual property that gives its owner the legal right to exclude others from making, using, selling, and importing an invention for a limited period of years, in exchange for publishing an enabling public disclosure of the invention \cite{bagley2003patent}. For the analysis, the World Bank Open Data \cite{azevedo2019wbopendata}, a database that collects important indicators of the economy, education, health care for all countries in the world, was used. 

In 1985, Cuba registered 214 international patents, while in 2018 this number was only 29. Although in Latin America there is a steady trend regarding the creation of patents, countries such as Peru, Ecuador, Brazil, and Mexico have shown a substantial annual increase (4-fold increase from 1980- to 2017). Cuba is the country with the major fall in the production of patents during the last 20 years. By 2018, Cuba registered in the World Intellectual Property Organization patents with the following companies: CIGB (3), CIM (3), TECNOAZUCAR (3), and one patent with HABANOS, S.A., CUBATABACO, CUBACAFE and others (\href{https://www.wipo.int/ipstats/en/statistics/country\_profile/profile.jsp?code=CU}{details here}). A similar pattern was experienced by Russia in the ’90s since the number of patents registered decreased from 39494 in 1992 to 16454 in 1998. Former soviet republics such as Romania and Bulgaria decreased from 3000 patents in 1985 to roughly 1000 in 2000). From 2008-2017, Spain also experienced a major decrease in the number of patents caused by the financial crisis. Biotechnology remains the only innovation sector in Cuba, led by two institutions: CIGB and CIM. 

\subsection{Scientific publications by research topic and institutions}

\begin{figure}[ht]
  \centering
  \includegraphics[width=15cm]{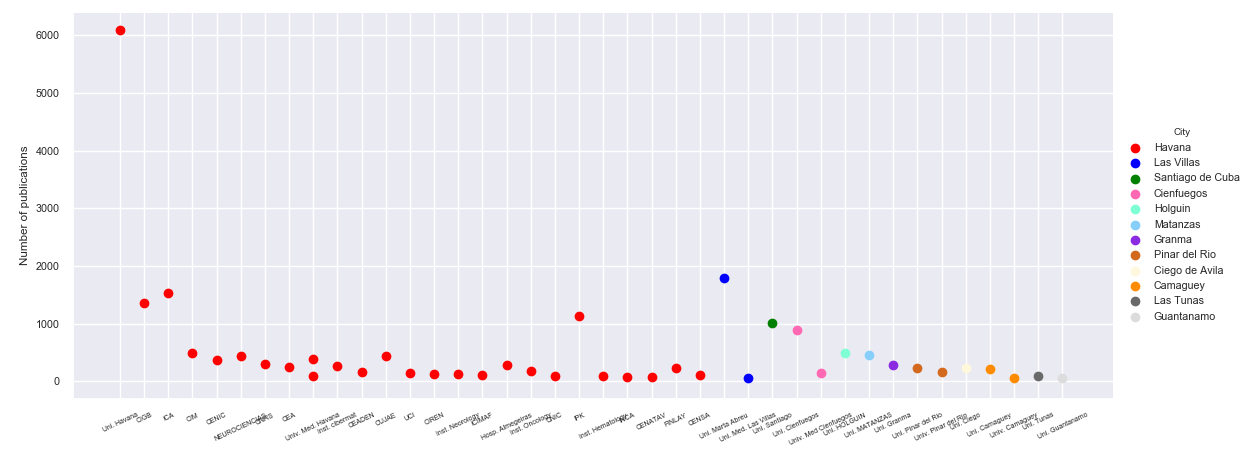}
  \caption{Number of scientific publications between 1970 and 2019 by research institutions and provinces in Cuba.}
  \label{fig:fig4}
\end{figure}

 Research is not homogeneous in Cuba. Major differences occur among provinces; type of institutions; and research topics. In Cuba (Figure \ref{fig:fig4}) up to 70\% of the scientific publications (1970 - 2019) were published by a limited number of  institutions (11). These are, University of Havana (6089), Marta Abreu University of Las Villas (1794), Institute of Animal Science (1536), Center for Genetic Engineering and Biotechnology (1356), Institute of Tropical Medicine "Pedro Kourí" (1127), Santiago of Cuba University (1020), Cienfuegos University (890), Cuban Academy of Science (848), Center for Molecular Immunology (493), Holguin University (492), Matanzas University (451), Institute of Neurosciences (440). 

Interestingly, 70\% of the  published Cuban scientific article are produced from Havana’s institutions. In addition, more than 30\% of the national scientific output is produced in biomedical research institutions. On average, research institutions in Cuba publish 10 manuscripts per year whereas more than 20 institutions publish only 5 manuscripts per year. These numbers suggest that most of the researchers in universities and other institutions do not produce publications within a year. The University of Havana and Marta Abreu (Las Villas) shows major differences as compared with other Cuban universities.

\begin{figure*}[ht]
\begin{tabular}{cc}
  \includegraphics[width=8cm]{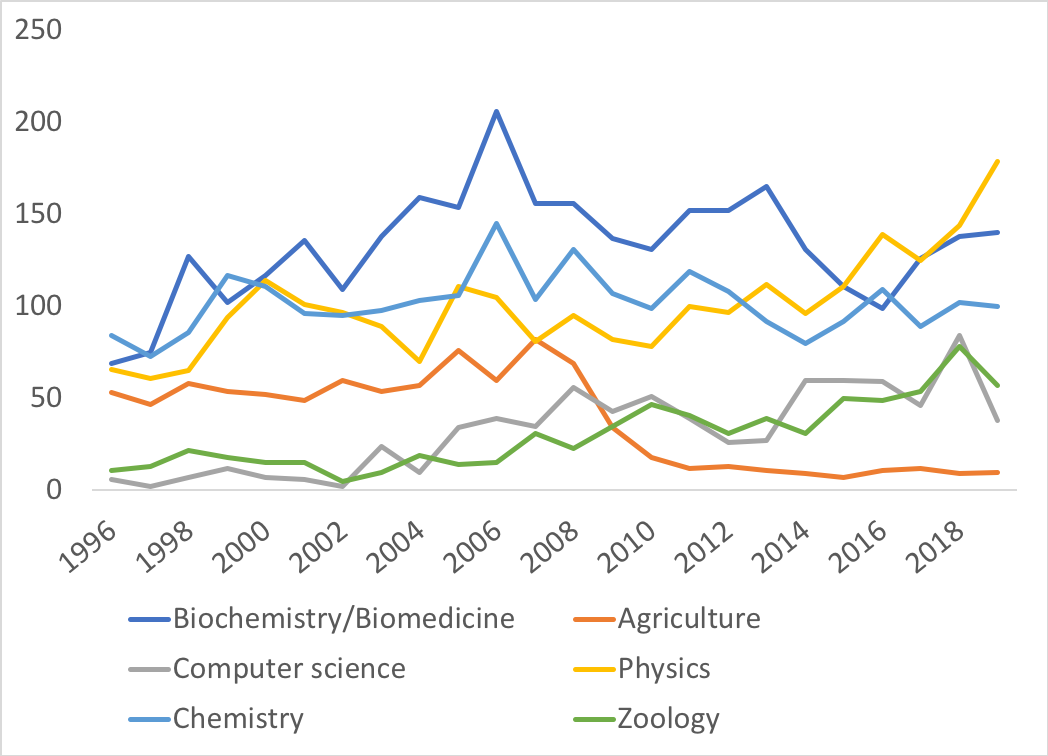} &   \includegraphics[width=8cm]{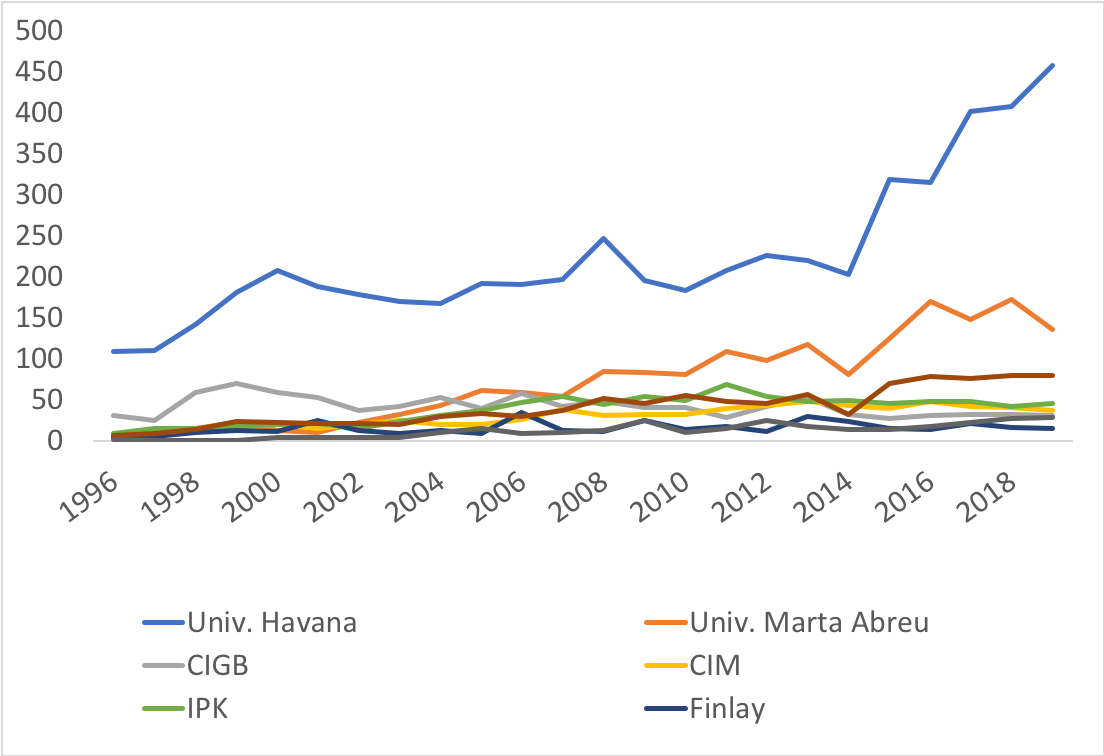} \\
  (a) Number of publications by research field. & (b) Number of publications by research institutions \\[6pt]
\end{tabular}
\caption{Number of publications between 1996 and 2019 by institutions and research topic}
\label{fig:fig5}
\end{figure*}

The number of publications on six major research topics from relevant institutions from 1996 to 2019, was also analised. Figure \ref{fig:fig5}a shows a dramatic fall in scientific output for agriculture research in Cuba and a slight decrease in biomedical and chemistry research. This can be confirmed at the level of institutions (Figure \ref{fig:fig5}b), where the Institute of Animal Science (ICA) decreased from 57 publications in 2002 to 10 publications in 2019. Physics, in contrast, continues the scientific output growth despite reports showing how the field has been affected in recent years by the lack of resources and emigration of young researchers \cite{marx2014physics}. Interestingly, computer science and software engineering research is not increasing the research output, despite major efforts of the country to develop new software companies and universities (e.g. University of \href{https://www.uci.cu/en}{Informatics Sciences}).  

\subsection{Knowledge networks and emigration: two variables to explain 
the Cuban scientific output patterns}

Economic factors are crucial to define the science productivity of a country and scientific growth. However, other factors such knowledge networks and emigration can profoundly affect the scientific output of a country and its institutions \cite{welch2008higher, huggins2012knowledge}. In 2014, Nature published a series of comment papers about the impact of several factors in science such as emigration and “brain circulation” \cite{Wiesel2014, Fraser2014}. Two different factors suggest Knowledge networks and emigration can be playing a major role in Cuban scientific output: 

\begin{itemize}
  \item The differences between some institutions like the University of Havana, CIGB, and University Martha Abreu with the rest of the institutions of the country, suggest that clusters of knowledge are created on these institutions and a proficient culture of research. While research institutions and companies has more resources to perform research, universities across the country should be receiving similar resources, then differences are mainly related to the personal, the culture of publications, and research topics. 

  \item The Cuban economic crisis caused by the end of the Soviet Union started in the early '90s. However, the major fall in the annual number of publication were observed after 2010, suggesting that other factors such as emigration have a major impact. This is especially true for universities.  

\end{itemize}

\begin{figure}[ht]
  \centering
  \includegraphics[width=10cm]{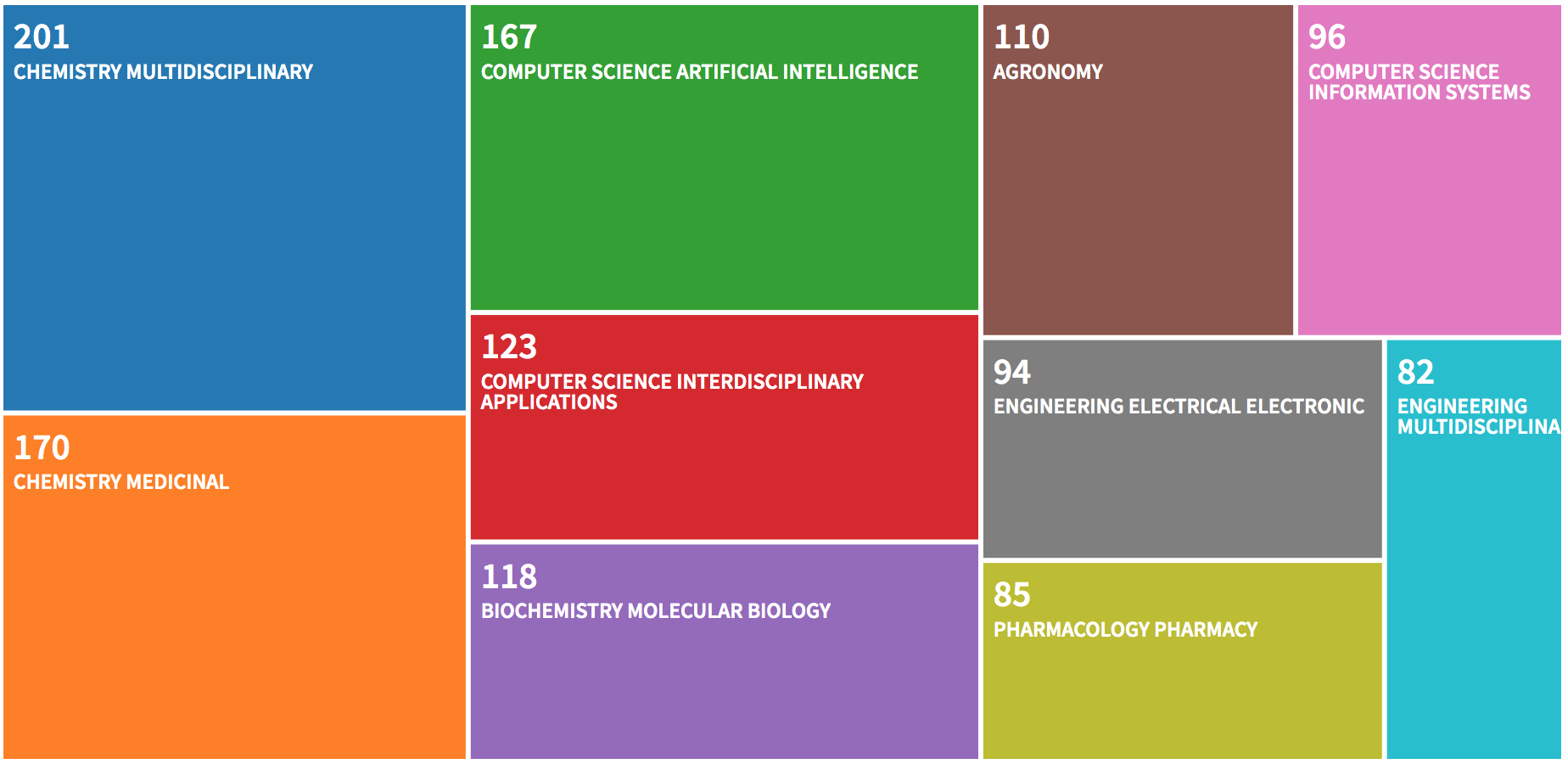}
  \caption{Clusters by topic of manuscripts from Marta Abreu (1970 - 2019)}
  \label{fig:fig6}
\end{figure}

\begin{figure}[ht]
  \centering
  \includegraphics[width=15cm]{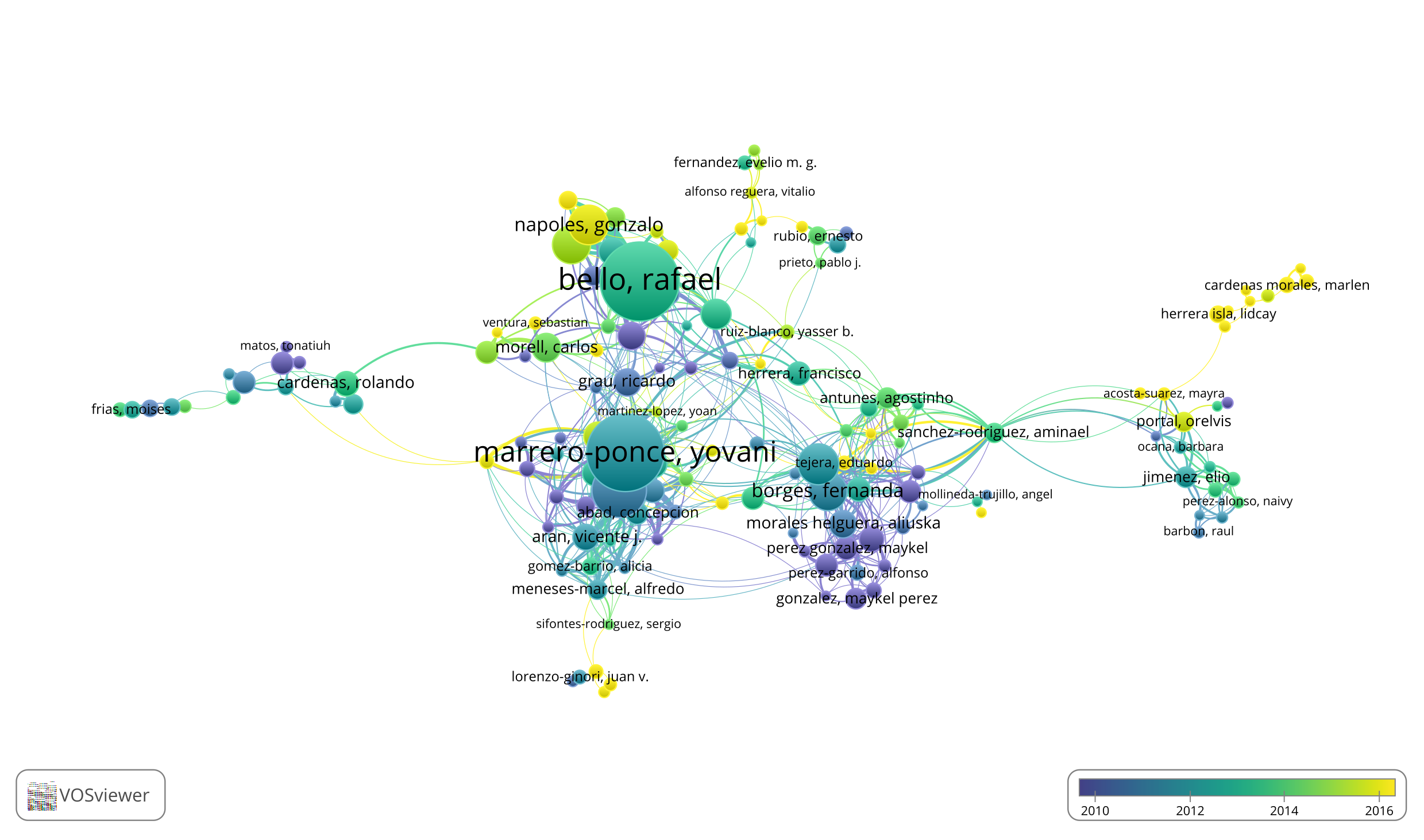}
  \caption{Authors network for all Marta Abreu publications in Web of Science (between 1970 - 2019) }
  \label{fig:fig7}
\end{figure}

\begin{figure}[ht]
  \centering
  \includegraphics[width=15cm]{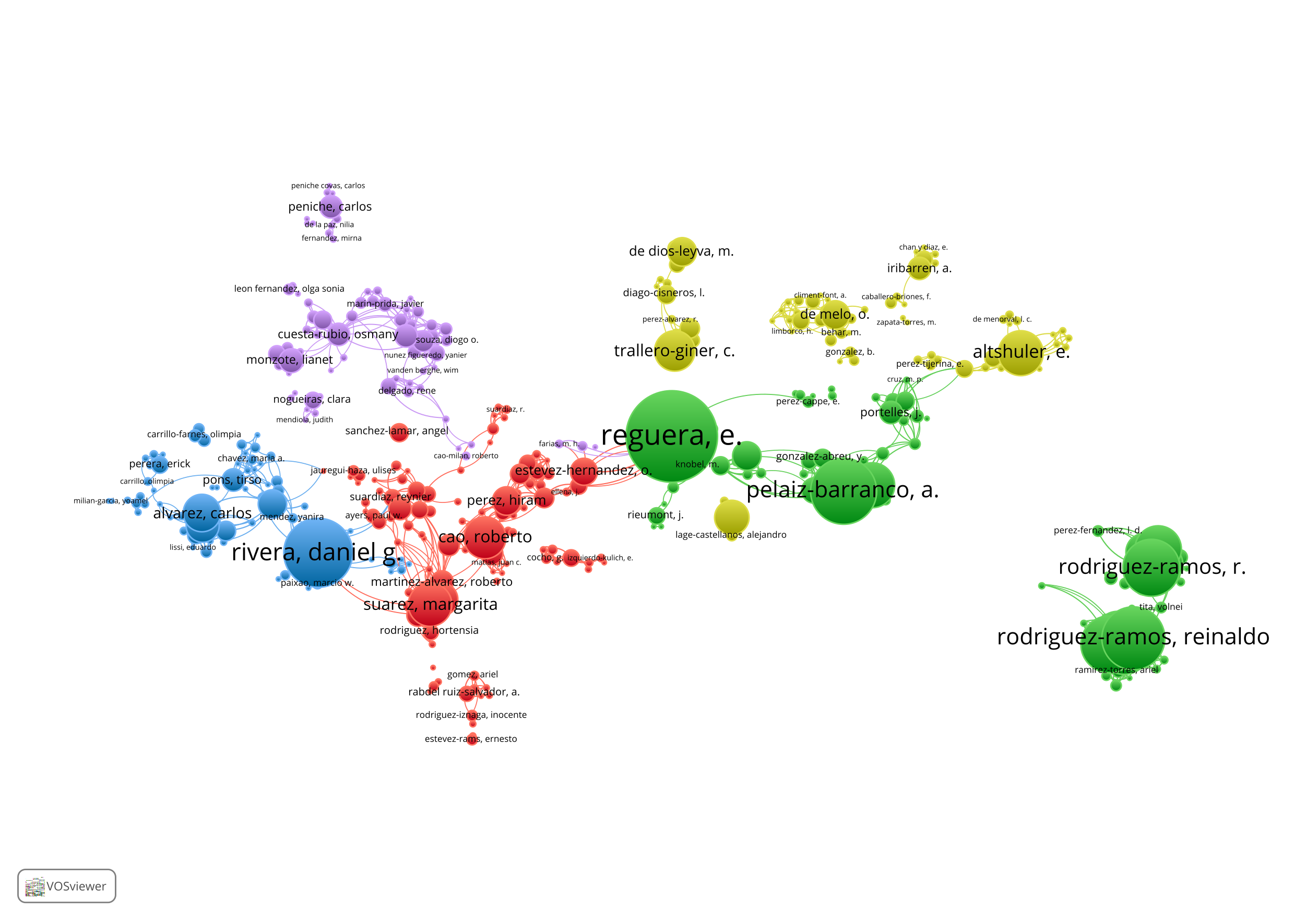}
  \caption{Authors network for all Havana University publications in Web of Science (between 2000 - 2019) }
  \label{fig:fig8}
\end{figure}

Knowledge networks are important because they build the bases for new scientific discoveries and it also help the transition across generations. At the same time, if the network are poorly connected they can be affected by different factors such as emigration, staff movements or changes in topics. The publications from the two most productive universities in the country, University Martha Abreu (Figure \ref{fig:fig7}) and University of Havana, were studied (Figure \ref{fig:fig8}).

Up to 1845 publications from Marta Abreu University were compiled and clustered by topic. Remarkably, (Figure \ref{fig:fig6}) shows that more than 60\% of the publications are in the field of computational chemistry and computer science. The development of these two fields can explain the major differences with other universities in the country. Interestingly, five authors have produced 20\% of all manuscripts and one author \href{https://scholar.google.com/citations?user=rsbUYyEAAAAJ}{Marrero-Ponce} has produced 5\% (100 manuscripts) of all the manuscripts of the University (Figure \ref{fig:fig7}). This pattern is knowledge network is extremely fragile to changes. For example, emigration or movement of some of the professors can heavily affect the scientific production. In fact, the analysis showed that some former  highly productive authors from this university are working abroad. This may explain the drop in the number of publications of this academic institution from 144(2019) to 173 (2018). 

In contrast, University of Havana shows more diverse and interconnected network with different hubs (Figure \ref{fig:fig8}). Strong networks of multiple researchers are present in the Physics, Chemistry and Biology faculties. In these cases, most of the authors are still in Cuba and that can explain why, unlike most instittions in Cuba, the scientific output of the University of Havana keeps growing annually (Figure \ref{fig:fig5}). 

\subsubsection{Emigration}

\begin{figure}[ht]
  \centering
  \includegraphics[width=15cm]{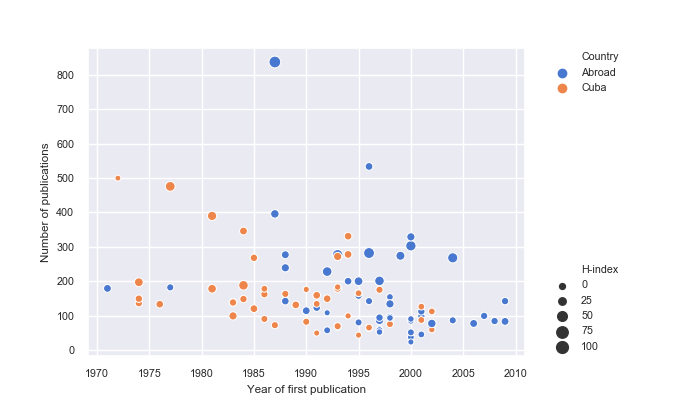}
  \caption{Cuban authors over 1000 citations. The x-axis corresponds to the year of the first publication, that y-axis to the number of publications (including thesis, etc), the size of the bubble to the Google Scholar H-index and the color of the bubble where the author is currently living (blue – abroad, orange – Cuba).}
  \label{fig:fig9}
\end{figure}

In order to explore how emigration of Cuban scientists have affected the scientific output of Cuban institutions, a list of the 100 Cuban authors living in the country or abroad with over 1000 citations in Google Scholar or ResearchGate, was curated (Figure \ref{fig:fig9}). 

Most of the authors that published their first scientific papers before 1995 (~48 years old) are living in Cuba, while most authors that published their first manuscript after 1995 are living abroad. Most of these young authors have a high h-index considering that their first paper was published around 1995. The 54 authors living abroad have published 8508 scientific manuscripts; 157 articles on average per researcher. These numbers are impressive if we consider that they represent 28\% of all publications on the Web of Science published by Cuban authors since 1970. Most of the researchers living abroad can be grouped into three major fields: biomedical research, computer science, and physics.

During the decades of the ‘60s and ‘70s, there was an increasing concern for the phenomenon of “brain drain”, which motivated its inclusion as a central subject in the agenda of discussions of different international organizations \cite{lowell2001migration}.  Recently, countries such as Brazil, Colombia, and Argentina introduced re-allocation programs for researchers and scientists. Under the RAICES program, for example, more than one thousand Argentinian scientist returns to the country after the economic crisis \cite{Fraser2014}. However, such programs do not exist in Cuban institutions, on the contrary, the number of young scientists emigrating increases every year. Examples such as Brazil evidenced that the emigration process can be reverted in science and are beneficial for the development because new ideas, technologies, and knowledge can emerge from reallocated scientists \cite{Fraser2014,Wiesel2014}. 

\section{Conclusions}

Recent bibliometrics studies have highlighted the increase in Cuban scientific output. However, the present analysis found some concerning patterns in the last two decades that should prompt attention. While the number of scientific publications is double every year for Cuban scientists, most of the countries of the region and the world are experiencing an increase of 5-fold in average. In addition, since 2014 the number of Cuban scientific publications is experiencing a decreasing pattern. The average number of publications per year for most Cuban universities is less than 5, except the University of Havana. Over 70\% of the articles are published from Havana’s institutions and more than 30\% of the scientific output is produced in biomedical research institutions. Results show a dramatic fall in scientific output for agriculture research and a slight decrease in biomedical and chemistry research in Cuba.

Regarding intellectual property, while most of the LA countries produce 4 times more patents in 2018 compared with 1980, in Cuba the number of patents has decreased from almost 300 per year in 1980 to 30 in 2018. Biotechnology remains the only innovation sector on the island, led by two institutions: CIGB and CIM. 

We finally studied the number of citations and manuscripts of the young scientist with more than 1000 GoogleScholar or ResearchGate citations. Most of the young Cuban authors (first published manuscript after 1995) live abroad. The 54 authors studied and living abroad have published 8508 scientific manuscripts; 157 articles on average per researcher. This prompts new challenges for Cuban science ahead to renew and revitalize their scientific community. This analysis can be used by policymakers, research institutions and the scientific community to assess and predict the Cuban science for the future, one of the more valuable treasures the Cubans have built in the last 50 years.

\section{Acknowledgements}

Thanks to Amilcar Perez-Riverol for reviewing the manuscript and  valuable feedback. 

\bibliographystyle{unsrt}  
\bibliography{references}  

\end{document}